\documentclass[prl,aps,showpacs,onecolumn,floatfix]{revtex4}
\usepackage{epsfig} \usepackage{graphics} \usepackage{bm}
\usepackage{amssymb}
\usepackage{graphicx}
\begin{document}

\preprint{Lebed-JP-2}

\title{Breakdown of the equivalence between active gravitational
mass and energy for a quantum body}

\author{Andrei G. Lebed$^*$}

\affiliation{Department of Physics, University of Arizona, 1118 E.
4-th Street, Tucson, Arizona 85721, USA; \
lebed@physics.arizona.edu}

\begin{abstract}
We determine active gravitational mass operator of the simplest
composite quantum body - a hydrogen atom - within the
semiclassical approach to the Einstein equation for a
gravitational field. We show that the expectation value of the
mass is equivalent to energy for stationary quantum states. On the
other hand, it occurs that, for quantum superpositions of
stationary states with constant expectation values of energy, the
expectation values of the gravitational mass exhibit
time-dependent oscillations. This breaks the equivalence between
active gravitational mass and energy and can be observed as a
macroscopic effect for a macroscopic ensemble of coherent quantum
states of the atoms. The corresponding experiment could be the
first direct observation of quantum effects in General Relativity.
\end{abstract}


\maketitle


\section{1. Introduction}
The notions of active and passive gravitational masses for a
classical composite body are not trivial and have been discussed
in recent literature by K. Nordtvedt [1] and S. Carlip [2]. In
particular, they have stressed that gravitational field is coupled
with the following combination: $3K+2U$, where $K$ is kinetic and
$U$ is potential energies of a composite body. They have revealed
an important role of the classical virial theorem, which states
that averaged over time value of $\bigl< 2K+U \bigl>_t = 0$ and
thus guarantees that the averaged over time gravitational masses
are equivalent to the total energy,
\begin{equation}
\bigl< m^g \bigl>_t = m_1 + m_2 + \bigl<3K+2U\bigl>_t/c^2 = m_1 + m_2 + \bigl<K+U\bigl>_t
/c^2 = E/c^2.
\end{equation}
The author of this article has very recently considered quantum
case of the simplest composite quantum body - a hydrogen atom
[3-7]. He has shown that the quantum virial theorem [8] is
responsible for the fact that the expectation values of active and
passive gravitational masses of the atom are equivalent to energy
for stationary quantum states. On the other hand, he has found
important breakdowns of the above-mentioned equivalence for two
cases: (a) for quantum superpositions of stationary states, (b)
for stationary quantum states due to quantum fluctuations. In
Refs.[3-7], there have suggested two different idealized
experiments to detect the above-mentioned breakdowns of the
Equivalence Principle [9]. If such experiments are done they will
be the first direct observations of quantum effects in General
Relativity.

\section{2. Goal}
The goal of this paper is to study a quantum problem of active
gravitational mass of a composite body in semiclassical theory of
gravity [10]. Below, we consider the simplest composite quantum
body - a hydrogen atom. We obtain and discuss the following two
main results. The first one is that the expectation value of the
mass is equivalent to energy for stationary quantum states due to
the quantum virial theorem [8]. The second result is the breakdown
of the above mentioned equivalence for a macroscopic coherent
ensemble of quantum superpositions of stationary states. In
particular, we show that the expectation value of active
gravitational mass is time dependent value for superpositions of
stationary quantum states even in the case, where the expectation
value of energy is constant. We also discuss possible experiment
to discover this breakdown of the Equivalence Principle.

\section{3. Active gravitational mass in classical physics}
In this section, we determine active gravitational mass of a
hydrogen atom, provided that we consider its classical model. More
precisely, below we consider light negatively charged particle
exhibiting a bound motion in the Coulomb field of heavy positively
charged particle. Our task is to calculate contributions to the
mass from kinetic and potential energies of the light particle.

Let us write gravitational potential at large distances from the
atom, $R \gg r_B$, where $r_B$ is the the so-called Bohr radius
(i.e., effective "size" of a hydrogen atom). In accordance with
general theory of a weak gravitational field [9,11], the
gravitational potential can be written as
\begin{equation}
\phi(R,t)=-G \frac{m_e + m_p}{R}- G \int \frac{\Delta
T^{kin}_{\alpha \alpha}(t,{\bf r})+ \Delta T^{pot}_{\alpha
\alpha}(t,{\bf r})}{c^2R} d^3 {\bf r} ,
\end{equation}
where $\Delta T^{kin}_{\alpha \beta}(t,{\bf r})$ and $\Delta
T^{pot}_{\alpha \beta}(t,{\bf r})$ are contributions to
stress-energy tensor density, $T_{\alpha \beta}(t, {\bf r})$, due
to kinetic and the Coulomb potential energies, respectively, $m_e$
and $m_p$ are electron and proton bare masses. We point out that
in Eq.(2) we disregard all retardation effects. Therefore, in the
above-mentioned approximation, electron active gravitational mass
is equal to
\begin{equation}
m^g_a = m_e + \frac{1}{c^2} \int [\Delta T^{kin}_{\alpha
\alpha}(t,{\bf r}) + \Delta T^{pot}_{\alpha \alpha}(t,{\bf r})]
d^3{\bf r}.
\end{equation}

To evaluate $\Delta T^{kin}_{\alpha \alpha}(t, {\bf r})$, we make
use of the standard expression for stress-energy tensor density of
a moving relativistic point mass [9,11]:
\begin{equation}
T^{\alpha \beta}({\bf r},t) = \frac{m_e v^{\alpha}(t)
v^{\beta}(t)}{\sqrt{1-v^2(t)/c^2}} \ \delta^3[{\bf r}-{\bf
r}_e(t)],
\end{equation}
where $v^{\alpha}$ is a four-velocity and ${\bf r}_e$ is three
dimensional electron trajectory. As directly follows from Eq.(4),
\begin{equation}
\Delta T^{kin}_{\alpha \alpha}(t) = \int \Delta T^{kin}_{\alpha
\alpha}(t,{\bf r}) d^3{\bf r} = \frac{m_e [c^2
+v^2(t)]}{\sqrt{1-v^2(t)/c^2}} -m_ec^2.
\end{equation}
Calculation of the contribution from potential energy to stress
energy tensor is done by using the standard formula for stress
energy tensor of electromagnetic field [11],
\begin{equation}
T_{em}^{\mu \nu} = \frac{1}{4 \pi} [F^{\mu \alpha} F^{\nu}_{\
\alpha} - \frac{1}{4} \eta^{\mu \nu} F_{\alpha \beta} F^{\alpha
\beta}],
\end{equation}
where $\eta_{\alpha \beta}$ is the Minkowski metric, $F^{\alpha
\beta}$ is the so-called tensor of electromagnetic field [11].
Below, we use approximation, where we disregard magnetic field and
take into account only the Coulomb electrostatic field. In this
case, we can simplify Eq.(6) and obtain the following expression:
\begin{equation}
\Delta T^{pot}_{\alpha \alpha} (t) = \int \Delta T^{pot}_{\alpha
\alpha}(t,{\bf r}) d^3{\bf r} = -2\frac{e^2}{r(t)}.
\end{equation}

As it follows from Eqs.(5),(7), active electron gravitational mass
can written in the following way
\begin{equation}
m^g_a = \biggl[\frac{m_e c^2}{(1 -v^2/c^2)^{1/2}} - \frac{e^2}{r}
\biggl]/c^2 + \biggl[\frac{m_e v^2}{(1
-v^2/c^2)^{1/2}}-\frac{e^2}{r}\biggl]/c^2.
\end{equation}
We note that the first term in Eq.(8) is the expected total energy
contribution to the mass, whereas the second term is the so-called
relativistic virial one [8,12], which depends on time. Therefore,
in classical physics, active gravitational mass depends on time
too. Nevertheless, it is possible to introduce electron active
gravitational mass averaged over time. This procedure restores the
expected equivalence between active gravitational mass and energy:
\begin{equation}
m^g_a = \biggl<\frac{m_e c^2}{(1 -v^2/c^2)^{1/2}} - \frac{e^2}{r}
\biggl>_t/c^2 + \biggl<\frac{m_e v^2}{(1
-v^2/c^2)^{1/2}}-\frac{e^2}{r}\biggl>_t/c^2 = m_e + E/c^2,
\end{equation}
where the averaged over time virial term is zero due to the
classical virial theorem. Note that for non-relativistic particle
our Eqs.(8),(9) can be reduced to the results of Refs.[1,2]:
\begin{equation}
m^g_a = m_e + \biggl[\frac{m_e v^2}{2} - \frac{e^2}{r} \biggl]/c^2
+ \biggl[2 \frac{m_e v^2}{2}-\frac{e^2}{r}\biggl]/c^2
\end{equation}
and
\begin{equation}
m^g_a = m_e + \biggl<\frac{m_e v^2}{2} - \frac{e^2}{r}
\biggl>_t/c^2 + \biggl<2 \frac{m_e
v^2}{2}-\frac{e^2}{r}\biggl>_t/c^2 = m_e + E/c^2.
\end{equation}

\section{4. Gravitational mass in quantum physics}
In this section, we make use of semiclassical theory of gravity
[10], where gravitational field is not quantized but the matter is
quantized in the Einstein equation:
\begin{equation}
R_{\mu \nu} - \frac{1}{2}R g_{\mu \nu} = \frac{8 \pi G}{c^4}
\bigl<\hat T_{\mu \nu} \bigl> ,
\end{equation}
where $<\hat T_{\mu \nu}>$ stands for the expectation value of
quantum operator, corresponding to the stress energy tensor. To
this end, we need to rewrite Eq.(10) for electron active
gravitational mass using momentum, instead of velocity, and then
quantize it:
\begin{equation}
\hat m^g_a = m_e +\biggl(\frac{{\bf \hat
p}^2}{2m_e}-\frac{e^2}{r}\biggl)/c^2 + \biggl(2\frac{{\bf \hat
p}^2}{2m_e}-\frac{e^2}{r}\biggl)/c^2.
\end{equation}
As follows from Eq.(13), the expectation value of electron active
gravitational mass can be expressed as
\begin{equation}
<\hat m^g_a> = m_e +\biggl< \frac{{\bf \hat
p}^2}{2m_e}-\frac{e^2}{r}\biggl>/c^2 + \biggl<2\frac{{\bf \hat
p}^2}{2m_e}-\frac{e^2}{r}\biggl>/c^2,
\end{equation}
where third term is the virial one. Let us consider a macroscopic
ensemble of hydrogen atoms with each of them being in ground
state. In this case, the expectation value of the mass is
\begin{equation}
<\hat m^g_a> = m_e + \frac{E_1}{c^2},
\end{equation}
where the expectation value of the virial term in Eq.(14) is equal
to zero in stationary quantum states due to the quantum virial
theorem [8]. Thus, we make conclusion that, in stationary quantum
states, active gravitational mass of a composite quantum body is
equivalent to its energy.

Here, we consider the simplest quantum superposition of stationary
states in a hydrogen atom,
\begin{equation}
\Psi (r,t) = \frac{1}{\sqrt{2}} \bigl[ \Psi_1(r) \exp(-iE_1t) +
\Psi_2(r) \exp(-iE_2t) \bigl],
\end{equation}
where $\Psi_1(r)$ and $\Psi_2(r)$ are the ground state (1S) and
first excited state (2S), respectively. As directly follows from
(16), the superposition is characterized by the following constant
expectation value of energy:
\begin{equation}
<E> = (E_1+E_2)/2.
\end{equation}
Nevertheless, as it follows from (14) the expectation value of
electron active mass operator oscillates with time:
\begin{equation}
<\hat m^g_a> = m_e + \frac{E_1+E_2}{2 c^2} + \frac{V_{1,2}}{c^2}
\cos \biggl[ \frac{(E_1-E_2)t}{\hbar} \biggl],
\end{equation}
where $V_{1,2}$ is matrix element of the virial operator between
the above-mentioned two stationary quantum states. Note that these
time dependent oscillations directly demonstrate inequivalence
between the expectation values of active gravitational mass and
energy for superpositions of stationary quantum states. We stress
that such quantum mechanical oscillations are very general and are
not restricted by a hydrogen atom. To simplify the situation, in
the same way as in the previous section, we can introduce the
averaged over time expectation value of active gravitational mass,
which obeys the Einstein's equation:
\begin{equation}
<<\hat m^g_a>>_t = m_e + \frac{E_1+E_2}{2 c^2} = \biggl<
\frac{E}{c^2} \biggl>.
\end{equation}

\section{5. Suggested experiment}
In this short section, we discuss in brief an idealized
experiment, which, in principle, allows to observe oscillations of
the expectation value of active gravitational mass (18). By using
laser, it is possible to create a macroscopic ensemble of coherent
superpositions of electron stationary states in some gas. It is
important that they are characterized by a feature that each
molecule has the same phase difference between two wave function
components, $\tilde \Psi_1(r)$ and $\tilde \Psi_2(r)$. In this
case, the ensemble of atoms generates gravitational field, which
oscillates in time (18).

\section{Acknowledgments}

We are thankful to N.N. Bagmet (Lebed), V.A. Belinski, Steven
Carlip, Fulvio Melia, Douglas Singleton, and V.E. Zakharov for
fruitful and useful discussions. This work was partially supported
by the NSF under Grant DMR-1104512.

$^*$Also at: L.D. Landau Institute for Theoretical Physics, RAS,
2 Kosygina Street, Moscow 117334, Russia.


\end{document}